\documentclass[a4paper]{article}

\usepackage{graphicx,array}
\usepackage{amssymb,amsmath}
\usepackage[mathscr]{euscript}
\usepackage[sans]{dsfont}
\usepackage[all]{xy}

\newcommand{\Acal}{\mathcal{A}}

\newcommand{\Fcal}{\mathcal{F}}

\newcommand{\Lcal}{\mathcal{L}}
 \newcommand{\Mcal}{\mathcal{M}}

 \newcommand{\Rcal}{\mathcal{R}}

\newcommand{\rmd}{\mathrm{d}}
\newcommand{\rme}{\mathrm{e}}
\newcommand{\rmi}{\mathrm{i}}
\newcommand{\rmf}{\mathrm{f}}
\newcommand{\Real}{\mathbb{R}}
\newcommand{\Complex}{\mathbb{C}}
\newcommand{\Qbb}{\mathbb{Q}}
\newcommand{\Pbb}{\mathbb{P}}

\newcommand{\abs}[1]{\left\vert#1\right\vert}

\newcommand{\RE}{\operatorname{Re}}

\newcommand{\Tr}{\operatorname{Tr}}
\newcommand{\EE}{\operatorname{\mathbb{E}}}

\title{Feedback control of the squeezing of the fluorescence light}

\author{  Alberto Barchielli and Matteo Gregoratti \\
        Politecnico di Milano,  Dipartimento di Matematica,\\
         Piazza Leonardo da Vinci 32,  20133 Milano,  Italy\\
      and INFN (Istituto Nazionale di Fisica Nucleare)\\
    }

\begin{document}

\maketitle

\begin{abstract}
Among the formulations of the theory of quantum measurements in continuous time, quantum
trajectory theory is very suitable for the introduction of measurement based feedback and
closed loop control of quantum systems. In this paper we present such a construction in the
concrete case of a two-level atom stimulated by a coherent, monochromatic laser. In
particular, we show how fast feedback \`{a} la Wiseman and Milburn can be introduced in the
formulation of the theory. Then, the spectrum of the free fluorescence light is studied and
typical quantum phenomena, squeezing and sub-natural line-narrowing, are presented.
\end{abstract}

\section{Introduction}

The theory of continuous measurements and filtering for quantum systems \cite{Bel88,BarB91}
has opened the possibility of a consistent introduction of measurement based feedback in
continuous time and, so, of closed loop control of quantum systems. The formulation of
continuous measurement theory adopted in the paper is the one based on stochastic differential
equations, commonly known in quantum optics as quantum trajectory theory \cite{Carm08}.

Photo-detection theory in continuous time has been widely developed inside quantum
trajectories \cite{BarP96,ZolG97} and applied, in particular, to the fluorescence light
emitted by a single trapped two-level atom stimulated by a coherent monochromatic laser. As
well as various feedback schemes on the atom evolution, based on the outcoming photocurrent,
have been proposed \cite{WisM93}. However the introduction and the analysis of feedback have
been mainly focused on the control of the atom. The typical aim was to drive the atom to a
preassigned asymptotic state.

Here, we are interested not only in the atom, but also, and mainly, in the emitted light. Of
course, a detection of the emitted light can be seen as a direct observation of the emitted
light as well as an indirect observation of the emitting atom mediated by the fluorescence
light. Our aim is to employ control and feedback processes to enhance the squeezing properties
of the fluorescence light. The squeezing of the fluorescence light can be checked by homodyne
detection and spectral analysis of the output current. We consider the mathematical
description of photo-detection based on quantum trajectories, as it is suitable both to
consistently compute the homodyne spectrum of fluorescence light, and to introduce feedback
and control in the mathematical formulation. We study how the squeezing depends on the various
control parameters, how feedback mechanisms can be successfully introduced and used to enhance
the squeezing \cite{BarGL09,BarG09}. We consider only the Markovian feedback scheme \`a la
Wiseman-Milburn \cite{WisM93}, which is simple, but flexible enough to give a physically
interesting model of closed loop control. We also show that, under certain conditions,
feedback can produce an effect of line narrowing in the homodyne spectrum, similarly to what
happens to an atom stimulated by squeezed light.

Apart from the new results on enhancing squeezing by feedback and on line narrowing, there is
a conceptual interest in using quantum trajectory theory in deriving the homodyne spectrum.
The traditional approach to homodyne or heterodyne spectrum is to define, by some analogy with
the classical case, a suitable quantum correlation function for the outgoing electromagnetic
field. Then, the spectrum is defined to be the Fourier transform of such a correlation
function. Finally, the quantum fields are eliminated in favour of the atomic variables and the
``quantum regression theorem''  is used to get the final result. On the other side, quantum
trajectory theory is based on quantum measurement theory and gives the output of the
measurement in continuous time together with its distribution, consistently with principles of
quantum mechanics. As the output is a stochastic process, its spectrum can be rigorously
defined by probability theory by using its distribution. Moreover, thanks to the consistency
of quantum continuous measurement theory, even when the quantum fields have been already
eliminated (so that quantum regression theorem is implicitly already contained), the output
can be seen not only as an observation of the system, but also as the result of an observation
of the output field which mediates the measurement. To succeed in describing a quantum effect
such as squeezing of the fluorescence light by using quantum trajectory theory is to show that
such a theory, in spite of its ``classical flavour'', is fully quantum mechanical.

We consider a trapped two-level atom with Hilbert space $\Complex^2$. Denoting by $\vec\sigma$
the vector $(\sigma_x,\sigma_y,\sigma_z)$ of the Pauli matrices, let the free Hamiltonian of
the atom be $\omega_0\sigma_z/2$, with resonance frequency $\omega_0>0$, and let the lowering
and rising operators be $\sigma_-$ and $\sigma_+$. Let the eigenprojectors of $\sigma_z$ be
denoted by $P_+$ and $P_-$ and, for every angle $\phi$, let us introduce the unitary
selfadjoint operator
\begin{equation*}
\sigma_\phi = \rme^{\rmi\phi}\,\sigma_- + \rme^{-\rmi\phi}\,\sigma_+
= \cos\phi\,\sigma_x+\sin\phi\,\sigma_y.
\end{equation*}

\section{The dynamics}
We start by giving the theory in the case of no feedback and then we show how to introduce it,
according to the scheme of \cite{WisM93}.
\subsection{No feedback}
We admit an open Markovian evolution for the atom, subjected to `dephasing' effects and to
interactions both with a thermal bath and with the electromagnetic field, via absorption and
emission of photons. The atom is stimulated by a coherent monochromatic laser. We denote by
$\gamma>0$ the natural linewidth of the atom, by $\Omega\geq0$ the Rabi frequency, by
$\omega>0$ the frequency of the stimulating laser and by $\Delta\omega =\omega_0-\omega$ the
detuning. Other parameters are the intensities of the dephasing and thermal effects, $k_\rmd
\geq 0$ and $\overline n\geq 0$. The state $\eta_t$ is governed by a Master equation, which
turns out to be time-homogeneous in the rotating frame (which we adopt here and in the rest of
the paper):
\[
\frac{\rmd\ }{\rmd  t}\, \eta_t= \Lcal\eta_t\,,
\]
where
\begin{multline*}
\Lcal\rho = -\rmi\left[\frac{\Delta\omega}{2}\,\sigma_z +
\frac{\Omega}{2}\,\sigma_x\;,\;\rho\right] + \gamma k_\rmd
\left(\sigma_z\,\rho\,\sigma_z-\rho\right)
\\ {}+
\gamma\overline{n}\left(\sigma_+\,\rho\,\sigma_--\frac{1}{2}\left\{P_-\;,\;\rho\right\}\right)
+\gamma(\overline{n}+1) \left(\sigma_-\,\rho\,\sigma_+-\frac{1}{2}
\left\{P_+\;,\;\rho\right\}\right).
\end{multline*}
Note that the intensity of the laser enters only in the effective Hamiltonian through the Rabi
frequency $\Omega$.

The measuring apparatus is made by two homodyne detectors. Part of the emitted light reaches
the detectors and part is lost in the free space. The fraction of light detected by one of the
detectors depends on its efficiency, on the spanned solid angle and can eventually be enhanced
by using a focussing mirror. So, the fluorescence light is divided in three parts according to
the direction of propagation: we call \textit{side channel $k$} ($k=1,2$) the directions
reaching the detector $k$, and \textit{forward channel} those of the lost light. The
stimulating laser is well collimated in such a way that it does not hit the detectors; so, we
can say that it acts in the forward channel. We denote the effective fractions of light
emitted in the forward and in the two side channels by $|\alpha_0|^2$, $|\alpha_1|^2$,
$|\alpha_2|^2$, respectively; obviously, $|\alpha_0|^2+|\alpha_1|^2+|\alpha_2|^2=1$. For
$k=1,2$, we can also interpret $|\alpha_k|^2$ as the total efficiency of the detector $k$.
Moreover, the initial phase of the local oscillator in each detector is denoted by
$\vartheta_k$ and it is included in the parameter $\alpha_k\in\mathbb{C}$ by setting
$\vartheta_k=\arg\alpha_k$. To change $\vartheta_k$ means to change the measuring apparatus.

We can condition the evolution of the atom on the continuous monitoring of the photocurrents.
The two homodyne photocurrents $I_1$ and $I_2$ and the conditional state of the atom $\rho_t$
(a posteriori state) are stochastic processes whose distributions depend on the initial state
of the atom and on the parameters introduced so far. In particular the atom has still a
Markovian evolution, even if stochastic. Let us introduce the latter by means of the
\emph{linear} stochastic Master equation
\begin{equation}\label{posterior}
\rmd\sigma_t= \Lcal\sigma_t\,\rmd t + \sqrt\gamma\sum_{j=1}^2\Rcal[\alpha_j\sigma_-]
\sigma_t\,\rmd W_j(t),
\end{equation}
where, for every matrix $a$, the superoperator $\Rcal[a]$ is
\begin{equation*}
\Rcal[a]\rho=a\,\rho + \rho\,a^*,
\end{equation*}
and where $W_1$ and $W_2$ are two independent standard Wiener processes in some reference
probability space$(\Omega,\Fcal,\Qbb)$. In all these equations the initial condition is a
statistical operator $\varrho_0=\eta_0=\sigma_0$ and the solution $\sigma_t$ of
\eqref{posterior} is a positive operator valued stochastic process.

Then, the homodyne photocurrents are the (generalised) stochastic processes
$I_j(t)=\dot{W}_j(t)$, while the \emph{a posteriori state} is
\[
\rho_t:=\frac{\sigma_t}{p_t}\,, \qquad \text{with} \quad p_t=\Tr\{\sigma_t\}.
\]
The important point is that the physical distribution of the homodyne currents and the a
posteriori states in the time interval $[0,T]$ is given by the new probability $\Pbb_T(\rmd
\omega)= p_T(\omega)\Qbb(\rmd \omega)$: the quantity $p_T=\Tr\{\sigma_T\}$ is the density of
the \emph{physical probability} with respect to the reference probability. By the fact that
$p_t$, $t\geq 0$, turns out to be a $\Qbb$-martingale, we have that the physical probabilities
do not depend on the final time $T$.

Of course, we could switch on the detectors, but decide to ignore the outputs. This should not
modify the evolution of the atom and, indeed, for every $t\leq T$,
\[
\eta_t=\int_\Omega\sigma_t(\omega)\,\Qbb(\rmd\omega)=\int_\Omega\rho_t(\omega)\,\Pbb_T(\rmd\omega).
\]

Thanks to the linearity of \eqref{posterior}, we can introduce the stochastic evolution map
(or propagator) $\Acal(t,s)$, satisfying
\begin{equation*}
\Acal(t+\rmd t,t) - \mathrm{Id} = \Lcal\rmd t + \sqrt\gamma\sum_{j=1}^2\Rcal[\alpha_j\sigma_-]\rmd W_j(t),
\end{equation*}
so that $\sigma_t=\Acal(t,0)\,\sigma_0$. Here, $\mathrm{Id}$ is the identity map on the space
of $2\times 2$ complex matrices.

\subsection{Introduction of the feedback}
We introduce a feedback scheme based on $I_1$, in order to modify the properties of the
fluorescence light in channel 2. We check such properties by analysing the properties of
$I_2$, but, of course, we could remove the second homodyne detector and employ the light
emitted in channel 2 for other purposes. Let us call it \emph{free} light. Our scheme is
summarised by the following picture, where ``h.\ det.'' means ``homodyne detector'':
\begin{equation*}
\xymatrix{
& & & &
\\
*+[F]{\txt{\begin{scriptsize}h. det.\end{scriptsize}}} \ar[u]^>>>{I_2(t)}& &
*+[F-:<3pt>]{\txt{\begin{scriptsize}atom\end{scriptsize}}}
\ar@{~>}[u]^>>>{\text{forward}}_>>>{\text{channel}}\ar@{~>}[rr]^{\text{side}}_{\text{channel
1}} \ar@{~>}[ll]_{\text{side}}^{\text{channel 2}} & & *+[F]{\txt{\begin{scriptsize}h.
det.\end{scriptsize}}} \ar@/^1pc/[dll]^<<<<<<<{I_1(t)}
\\
&& *+[F]{\txt{\begin{scriptsize}electromodulator\end{scriptsize}}} \ar@{~>}[u]& &
\\
&& \ar@{~} [u]_<<<{\text{laser}} & &}
\end{equation*}
Assuming instantaneous feedback, the amplitude of the stimulating laser is modified by adding
a term proportional to $I_1$, with the same frequency $\omega$ and with initial phase possibly
different from that of the original laser. Let this phase difference be $\varphi\in[0,\pi)$.
As the laser intensity appears only in the Hamiltonian part of the evolution
\eqref{posterior}, the effect of the feedback is to give rise to a new Hamiltonian term
$I_1(t)\,M$ with $ M=c\sqrt{\gamma}  \,\sigma_\varphi$, where $c\in \Real$ controls the
intensity of the feedback. Let us deduce the modified evolution equation. By defining the map
$ \Mcal\rho=-\rmi[M,\rho]$, the contribution of the feedback to the propagator in an
infinitesimal interval turns out to be
\[
\rme^{\Mcal\, \rmd W_1(t)}=\mathrm{Id}+\Mcal \, \rmd W_1(t)+ \frac 1 2 \, \Mcal^2 \rmd t.
\]
Taking into account that the feedback must act after the signal is produced, the new
infinitesimal propagator is
\begin{equation*}
\Acal_{\rmf}(t+\rmd t,t) = \rme^{\Mcal\,\rmd W_1(t)}\circ\Acal(t+\rmd t,t).
\end{equation*}
By the Ito rules we obtain
\begin{multline*}
\Acal_{\rmf}(t+\rmd t,t) - \mathrm{Id} = \Lcal_{\rmf}\rmd t
\\ {}+ \sqrt\gamma\Rcal[\alpha_1\,\sigma_--\rmi c\,\sigma_\varphi]\,\rmd W_1(t) +
\sqrt\gamma\Rcal[\alpha_2\,\sigma_-]\,\rmd W_2(t),
\end{multline*}
where
\begin{eqnarray*}
\Lcal_{\rmf}\rho &=& -\rmi\left[\frac{\Delta\omega_c}{2}\,\sigma_z +
\frac{\Omega}{2}\,\sigma_x\;,\;\rho\right] + \gamma k_\rmd
\left(\sigma_z\,\rho\,\sigma_z-\rho\right)
\\ &+&
\gamma\overline{n}\left(\sigma_+\,\rho\,\sigma_--\frac{1}{2}\left\{P_-\;,\;\rho\right\}\right)
\\
&+&\gamma(\overline{n}+1-|\alpha_1|^2) \left(\sigma_-\,\rho\,\sigma_+-\frac{1}{2}
\left\{P_+\;,\;\rho\right\}\right)
\\ &+&\gamma(\alpha_1\,\sigma_--\rmi
c\,\sigma_\varphi)\,\rho\,(\overline{\alpha}_1\,\sigma_++\rmi c\,\sigma_\varphi)
\\ &-&
\frac{\gamma}{2}\left\{\Big(|\alpha_1|^2-2c|\alpha_1|\sin(\vartheta_1-\varphi)\Big)P_+
+c^2\;,\;\rho\right\},
\end{eqnarray*}
\[
\Delta\omega_c=\Delta\omega+c\,\gamma\,|\alpha_1|\cos(\vartheta_1-\varphi)\in\Real.
\]
Thus, thanks to the features of the particular feedback introduced, we get another linear
stochastic Master equation
\begin{equation}
\rmd\sigma_t= \Lcal_{\rmf}\sigma_t\,\rmd t + \sqrt\gamma\Rcal[\alpha_1\,\sigma_--\rmi
c\,\sigma_\varphi]\sigma_t\,\rmd W_1(t)
+ \sqrt\gamma\Rcal[\alpha_2\,\sigma_-]\sigma_t\,\rmd W_2(t).
\end{equation}

By Girsanov theorem, a fundamental result of stochastic calculus, it is possible to prove that
under the physical probability the output homodyne currents can be written as
\[
I_j(t)\rmd t= \rmd W_j(t)= \rmd B_j(t)+v_j(t)\rmd t,
\]
where $B_1(t)$, $B_2(t)$ ($t\in[0,T]$) are two independent standard Wiener processes under
$\Pbb_T$ and
\[
v_1(t)=2 \sqrt{\gamma} \RE \Tr \left\{\left(\alpha_1\sigma_--\rmi
c\sigma_\varphi\right)\rho_t\right\},
\qquad
v_2(t)=2 \sqrt{\gamma} \RE \left(\alpha_2\Tr \left\{\sigma_-\rho_t\right\}\right).
\]
Then the mean function of $I_2$ is
\[
\EE_{\Pbb_T}[I_2(t)] = v_2(t).
\]
Moreover, by using techniques based on characteristic functionals, explicit expressions for
the higher moments of the output have been obtained \cite{BarG09}. In particular, the
auto-correlation of $I_2$ is
\begin{multline}\label{2mom}
\EE_{\Pbb_T}[I_2(t)\,I_2(s)] = \delta(t-s) + \gamma \,
\Tr\Big\{\mathcal{R}[\alpha_2\sigma_-]\\ {}\circ\rme^{\abs{t-s}\mathcal{L}_f}
\circ\mathcal{R}[\alpha_2\sigma_-]\circ\rme^{\min\{s,t\}\mathcal{L}_f}\,\varrho_0\Big\}.
\end{multline}
The expression above is similar to what is obtained in traditional approaches through the
quantum regression theorem; the general theory of continuous measurements guarantees that it
is a positive definite function of $s$ and $t$ as it must be for an auto-correlation function.

\section{The spectrum of the free light in channel 2}

Let us consider now the spectrum of the light in channel 2, the fluorescence light detected
but not involved in the feedback loop. The homodyne current $I_2(t)$ is a stochastic process
and it can be proved that it is asymptotically stationary. Then, the spectrum $S_2(\mu)$ is
the Fourier transform of the autocorrelation function for long times. This classical
definition can be recast in the following way:
\[
S_2(\mu)=S_{\mathrm{el}}(\mu)+S_{\mathrm{inel}}(\mu),
\]
\[
S_{\mathrm{el}}(\mu)=\lim_{T\to+\infty}\frac{1}{T}\abs{\EE_{\Pbb_T}\left[\int_0^T\rme^{\rmi \mu
s}\,I_2(s)\,\rmd s\right]}^2 ,
\]
\begin{multline}\label{spectrum}
S_{\mathrm{inel}}(\mu)=\lim_{T\to+\infty}\frac{1}{T}\biggl\{\EE_{\Pbb_T}\left[\abs{\int_0^T\rme^{\rmi\mu
s}\,I_2(s)\,\rmd s}^2\right] \\ {} -\abs{\EE_{\Pbb_T}\left[\int_0^T\rme^{\rmi \mu
s}\,I_2(s)\,\rmd s\right]}^2\biggr\} \geq 0,
\end{multline}
where the spectrum is decomposed in the \emph{elastic} and \emph{inelastic} parts. By explicit
computations on can see that $S_{\mathrm{el}}(\mu)$ is proportional to a Dirac delta centred
in zero (as we are working in the rotating frame, $\mu=0$ corresponds to resonance with the
frequency of the stimulating laser). Instead, $S_{\mathrm{inel}}(\mu)$ is the limit of the
normalised variance of the Fourier transform of the photocurrent $I_2$. An explicit expression
can be obtained from the expression of the second moments \eqref{2mom}. The asymptotic
behaviour of the atomic a priori state $\eta_t=\rme^{\Lcal_ft}\varrho_0$ ensures that the
limit defining the spectrum exists and that it is independent of the initial state $\varrho_0$
of the atom. We get
\begin{equation}\label{spectrum2}
S_{\mathrm{inel}}(\mu)=1+2\gamma|\alpha_2|^2\,\vec{s}\cdot\left(\frac{A}{A^2+\mu^2}\,\vec{t}\right),
\end{equation}
where $\vec{s}=\begin{pmatrix}\cos\vartheta_2,&\sin\vartheta_2,&0\end{pmatrix}$, the matrix
$A$ has matrix elements
\[
a_{13}=a_{31}=0, \qquad a_{23}=-a_{32}=\Omega,
\]
\[
a_{11}=\gamma\Big(\frac{1}{2}+\overline{n}+2k_\rmd  + 2c|\alpha_1|\cos\vartheta_1
\sin\varphi+2c^2\sin^2\varphi\Big),
\]
\[
a_{12}=\Delta\omega_c - \gamma\Big(c|\alpha_1|\cos(\vartheta_1+\varphi)+c^2\sin2\varphi\Big),
\]
\[
a_{21}=-\Delta\omega_c - \gamma\Big(c|\alpha_1|\cos(\vartheta_1+\varphi)+c^2\sin2\varphi\Big),
\]
\[
a_{22}=\gamma\Big(\frac{1}{2}+\overline{n}+2k_\rmd \\ {} - 2c|\alpha_1|\sin\vartheta_1
\cos\varphi+2c^2\cos^2\varphi\Big),
\]
\[
a_{33}=\gamma\Big(1+2\overline{n} - 2c|\alpha_1|\sin(\vartheta_1-\varphi)+2c^2\Big),
\]
\[
\vec{t}=\Tr\Big\{\big(\rme^{\rmi\vartheta_2}\,\sigma_-\,\eta_\mathrm{eq} +
\rme^{-\rmi\vartheta_k}\,\eta_\mathrm{eq}\,\sigma_+
- \Tr\{\sigma_{\vartheta_2}\,\eta_\mathrm{eq}\}\,\eta_\mathrm{eq}\big)\,\vec\sigma\Big\}.
\]
and $\displaystyle \eta_\mathrm{eq}=\lim_{t\to +\infty}\rme^{\Lcal_ft}\varrho_0$.

In the case $|\alpha_2|=0$ we get $S_{\mathrm{inel}}(\mu)=1$, which is the spectrum of a pure
white noise. Indeed in this case no fluorescence light reaches the detector and we see only
the spectrum of the fluctuations of the local oscillator, interpreted as shot noise.

\begin{figure}[t]
\includegraphics*[scale=.35]{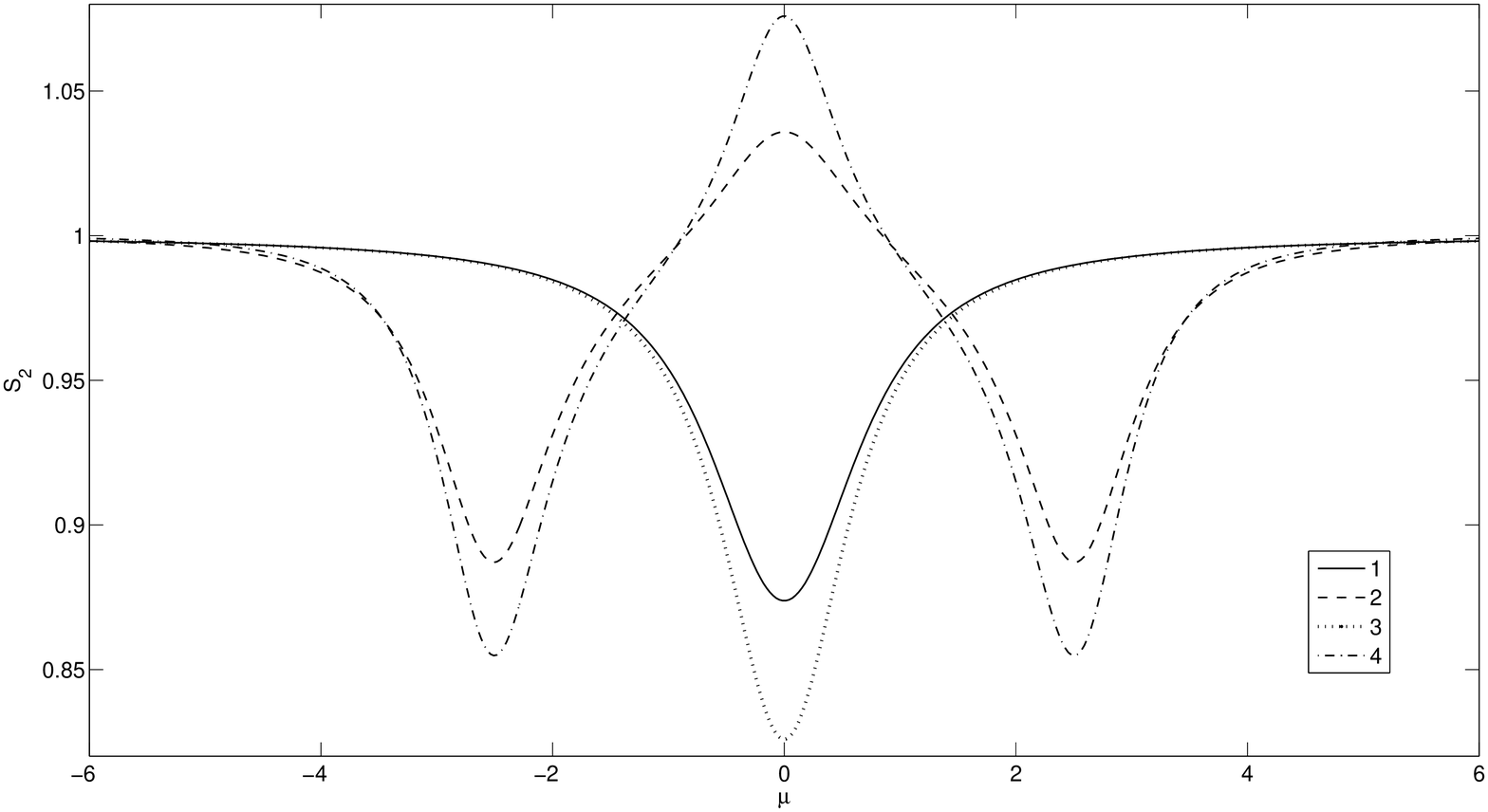}
\caption{$S_{\mathrm{inel}}(\mu)$ with $\gamma=1$ and: (1) $\Delta\omega=0$, $c=0$, $\Omega=0.2976$,
$\vartheta_2=-\pi/2$; (2) $\Delta\omega=1.8195$, $c=0$, $\Omega=1.7988$,
$\vartheta_2=-0.1438$; (3) $\Delta\omega=0$, $c=0.0896$, $\Omega=0.2698$,
$\vartheta_1=\pi/2$, $\vartheta_2=-\pi/2$, $\varphi=0$; (4)
$\Delta\omega=1.6920$, $c=0.1326$, $\Omega=1.9276$, $\vartheta_1=2.8168$,
$\vartheta_2=-0.0851$, $\varphi=1.2460$.}\label{fig1}
\end{figure}
\subsection{Squeezing}
In general it is possible to give a description of the output $I_2$ in terms of a measurement
of some Bose quantum fields, to show that the value of $S_{\mathrm{inel}}(\mu;\vartheta_2)$ is
the variance of a quadrature of the field in channel $2$, the value of
$S_{\mathrm{inel}}(\mu;\vartheta_2+\pi/2)$ is the variance of the conjugate quadrature
\cite{BarG08}. Then, the \emph{Heisenberg uncertainty relations} imply that
\[
S_{\mathrm{inel}}(\mu;\vartheta_2)\,S_{\mathrm{inel}}(\mu;\vartheta_2+\pi/2)\geq1.
\]
If $S_{\mathrm{inel}}(\mu;\vartheta_2)<1$ for some $\mu$ and $\theta_2$, the field is said to
be \emph{squeezed}.

Some typical spectra, showing well pronounced squeezing and produced by suitable choices of
the control parameters, are given in Figure \ref{fig1}. The position of the minimum can be
\emph{tuned} by a suitable choice of the control parameters; in Figure \ref{fig1} the minima
are in $\mu=0$ and in $\mu=\pm 2.5$, with and without feedback. The lines (1) and (2) are
without feedback and the lines (3) and (4) are with feedback. Here and in all the other
graphical examples we take $\abs{\alpha_1}^2=\abs{\alpha_2}^2=0.45$, $\overline n =0$ and
$k_\rmd=0$.

\begin{figure}[h]
\includegraphics*[scale=.35]{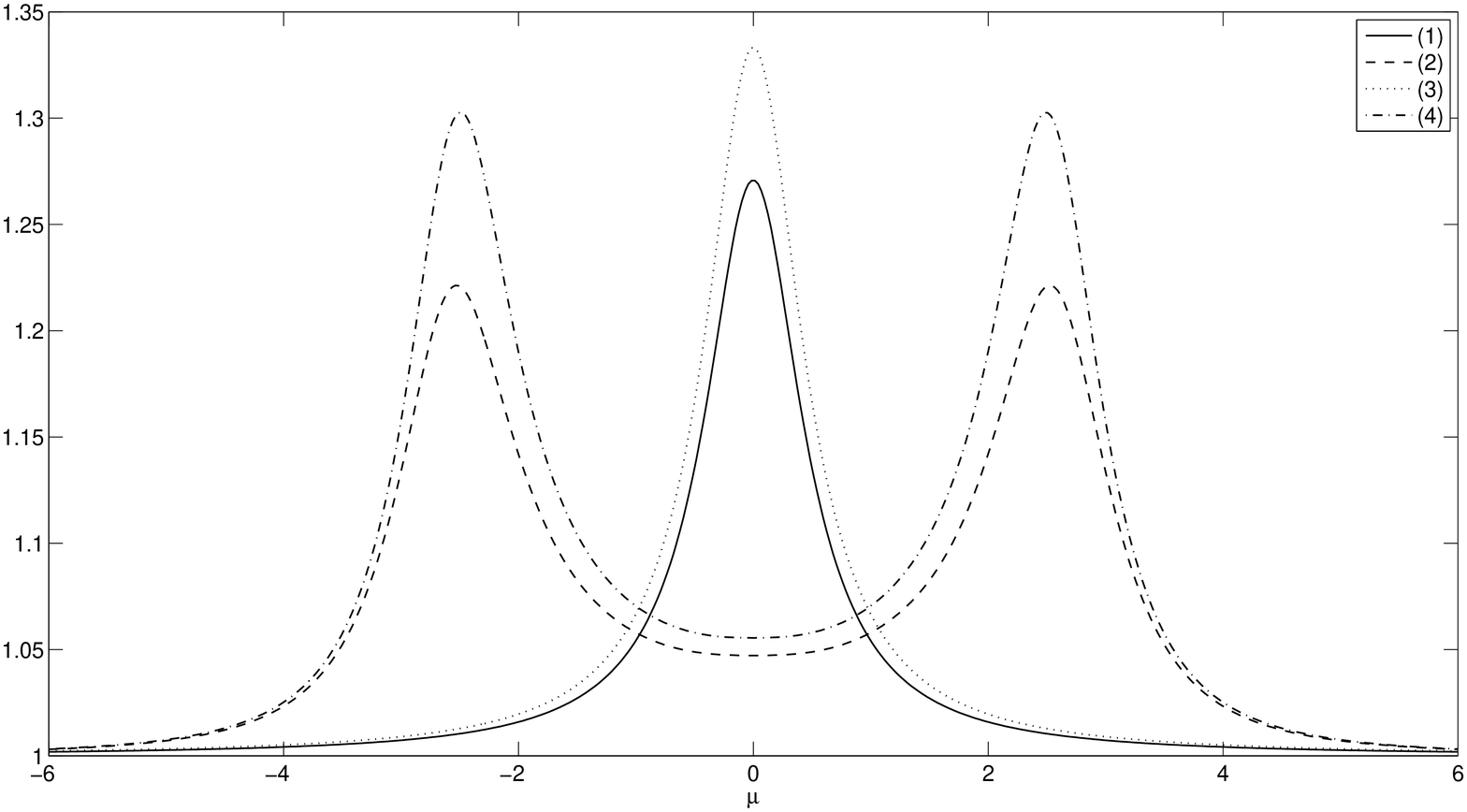}
\caption{$S_{\mathrm{inel}}(\mu)$ with all the parameters as in the previous figure, but
$\vartheta_2 \to \vartheta_2+\pi/2$.}\label{fig2}
\end{figure}

Note that the Heisenberg uncertainty relations give rise to peaks in the complementary field
quadratures, as shown in Figure \ref{fig2}.

\subsection{Line-narrowing}

By feedback control another quantum effect can be produced, the line-narrowing. After the
first observation of squeezing, Gardiner predicted that stimulating a two-level atom with
squeezed light would inhibit the phase decay of the atom. The squeezed light would break the
equality between the transverse decay rates for the two quadratures of the atom and one decay
rate could be made arbitrarily small, producing an observable narrow line in the spectrum of
the atomic fluorescence light. This was seen as a ``direct effect of squeezing'' and thus as a
measure of the squeezing of the incident light. Nevertheless, Wiseman showed that this atomic
line-narrowing is not only characteristic of squeezed light, but it can also be produced by
immersing a two-level atom in `in-loop squeezed' light. The difference is that in the Gardiner
case the other decay rate becomes larger, while in the Wiseman case it is left unchanged.

Now we can show that the same atomic line-narrowing can be obtained stimulating a two-level
atom even with non-squeezed light, that is with a coherent monochromatic laser in presence of
a (Wiseman-Milburn) feedback scheme based on the (homodyne) detection of the fluorescence
light. This effect is obtained within the model described up to now, in a different region of
the control parameters.

\begin{figure}[t]
\includegraphics*[scale=.35]{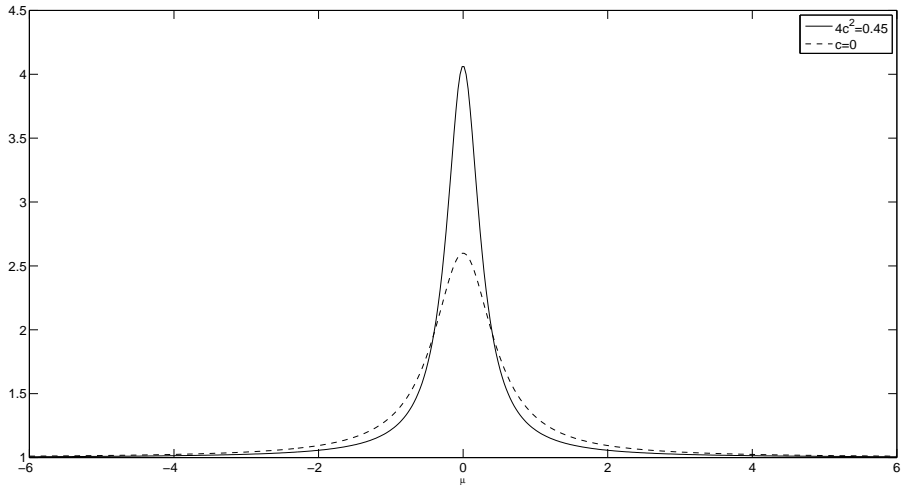}
\caption{
$S_{\mathrm{inel}}(\mu)$ for $c=\abs{\alpha_1}/2$ or $c=0$ and
$\Omega=2$, $p_1=p_2=0.45$, $\Delta \omega=0$, $\varphi=\frac \pi 2$, $\vartheta_1=\pi$,
$\vartheta_2=0$, $\gamma=1$.}\label{fig3}
\end{figure}

Let us take
\[
\varphi=\frac \pi 2 \,, \qquad \vartheta_1=\pi, \qquad \vartheta_2=0,
\qquad
\Delta \omega=0, \qquad \overline{n}=0, \qquad k_\rmd=0.
\]
Then, in the case of no feedback $c=0$ one obtains a spectrum which is a Lorentzian peak with
width $\gamma$:
\[ S_{\mathrm{inel}}(\mu)=1+\frac{2 \abs{\alpha_2}^2 \Omega^2}{
\gamma^2+2\Omega^2}\, \frac{\gamma^2}{\mu^2+\gamma^2/4}\,.
\]
Instead, with the optimal choice of the feedback, $ c=\abs{\alpha_1}/2$, we get a Lorentzian
of width $\left(1-\abs{\alpha_1}^2\right)\gamma$:
\[
S_{\mathrm{inel}}(\mu)=1+\frac{\abs{\alpha_2}^2\left(\frac{\abs{\alpha_1}^2}2\, \gamma^2+2
\Omega^2\right)} {\left(1-\frac{\abs{\alpha_1}^2}2\right) \gamma^2+2\Omega^2}
\frac{\left(1-\abs{\alpha_1}^2\right)
\gamma^2}{\mu^2+\left(1-\abs{\alpha_1}^2\right)^2\gamma^2/4}
\]
This is the sub-natural line-narrowing effect, shown in Figure \ref{fig3}.

\section{Conclusions}

The new results presented either are of general conceptual relevance, either clarify the
behaviour of fluorescence light and atom under coherent driving and feedback.

First of all we have shown how to introduce the spectrum in quantum trajectory theory through
its classical definition \eqref{spectrum}, without ad hoc quantum definitions, but in
agreement with the axiomatic structure of quantum measurement theory and with probability
theory. We have verified that this agreement with quantum measurement theory ensures
consistency with the existence of the traced out quantum field, so that Heisenberg uncertainty
relations hold for such spectra and the squeezing of the output field can be analysed.

We have shown how
feedback (at least in the form \`a la Wiseman-Milburn) modifies the spectrum of the free
fluorescence light and can enhance its squeezing. Many physical effects are introduced all
together: detuning, thermal and dephasing effects, not perfect detection efficiency, control.
The final formula for the spectrum \eqref{spectrum2} is given with all the parameters
introduced by these effects.

Finally, we have shown that feedback can produce line-narrowing in the free fluorescence
light, even if the atom is not illuminated by squeezed light, but only by coherent light.

\bibliographystyle{physcon}

\begin{thebibliography}{9}
\bibitem{BarB91}
    A. Barchielli, V. P. Belavkin, \textsl{Measurements continuous in time and a
    posteriori states in quantum mechanics}, J. Phys. A: Math. Gen. \textbf{24} (1991) 1495--1514;
    arXiv:quant-ph/0512189.

\bibitem{BarG08}
    A. Barchielli, M. Gregoratti, \textsl{ Quantum
    continual measurements: the spectrum of the output}. In J. C. Garc\'{\i}a, R. Quezada, S.
    B. Sontz (eds.), \textit{Quantum
    Probability and Related Topics}, Quantum Probability Series QP-PQ Vol.\ 23 (World
    Scientific, Singapore, 2008) pp.\ 63--76; arXiv:0802.1877v1 [quant-ph].

\bibitem{BarG09}
    A. Barchielli, M. Gregoratti, \textit{Quantum Trajectories and Measurements in
    Continuous Time: The Diffusive Case}, Lect.\ Notes Phys.\ \textbf{782} (Springer,
    Berlin, 2009); DOI 10.1007/978-3-642-01298-3.
\bibitem{BarGL09}
    A. Barchielli, M. Gregoratti, M. Licciardo, \textsl{
    Feedback control of the fluorescence light squeezing},  Europhysics Letters (EPL) \textbf{85}
    (2009) 14006;
    doi: 10.1209/0295-5075/85/14006; arXiv:0804.0085v1 [quant-ph].

\bibitem{BarP96} A. Barchielli, A. M. Paganoni, \textsl{
    Detection theory in quantum optics: Stochastic
    representation}, Quantum Semiclass. Opt. \textbf{8} (1996) 133--156.
\bibitem{Bel88} V. P. Belavkin, \textsl{Nondemolition measurements,
    nonlinear filtering and dynamic programming of quantum stochastic processes}. In A. Blaqui\`ere (ed.),
    \textit{Modelling and Control of Systems}, Lecture Notes in Control and Information
    Sciences \textbf{121} (Springer, Berlin, 1988) pp.~245--265.
\bibitem{Carm08} H. J. Carmichael, \textit{Statistical Methods in
    Quantum Optics 2. Non-Classical Fields} (Springer, Berlin, 2008).
\bibitem{ZolG97} P. Zoller, C. W. Gardiner, \textsl{
    Quantum noise in quantum optics: the stochastic Schr\"odinger equation}. In S. Reynaud, E. Giacobino,
    J. Zinn-Justin (eds.), \textit{Fluctuations quantiques, (Les Houches 1995)}
    (North-Holland, Amsterdam 1997) pp.~79–-136.
\bibitem{WisM93} H. M. Wiseman, G. J. Milburn, \textsl{
    Quantum theory of optical feedback via homodyne detection}, Phys. Rev. Lett. \textbf{70} (1993)
    548--551.

\end{thebibliography}

\end{document}